**Shot noise generated by graphene p-n junctions in the quantum Hall effect regime**

N. Kumada[1,2,*], F. D. Parmentier[2], H. Hibino[1], D. C. Glattli[2], and P. Roulleau[2]

[1]NTT Basic Research Laboratories, NTT Corporation, 3-1 Morinosato-Wakamiya, Atsugi 243-0198, Japan.

[2]Nanoelectronics Group, Service de Physique de l'Etat Condensé, IRAMIS/DSM (CNRS URA 2464), CEA Saclay, F-91191 Gif-sur-Yvette, France.

*Correspondence to: kumada.norio@lab.ntt.co.jp

Owing to a linear and gapless band structure and a tunability of the charge carrier type, graphene offers a unique system to investigate transport of Dirac Fermions at *p-n* junctions (PNJs), such as Klein tunnelling (ref: 1,2,3), Veselago lensing (ref: 4), and snake states (ref: 5, 6). In a magnetic field, combination of quantum Hall physics and the characteristic transport across PNJs leads to a fractionally quantized conductance (ref: 7, 8, 9) associated with the mixing of electron-like and hole-like modes and their subsequent partitioning (ref: 10, 11). The mixing and partitioning suggest that a PNJ could be used as an electronic beam-splitter. Here we report the shot noise study of the mode mixing process and demonstrate the crucial role of the PNJ length. For short PNJs, the amplitude of the noise is consistent with an electronic beam-splitter behavior, whereas, for longer PNJs, it is reduced by the energy relaxation. Remarkably, the relaxation length is much larger than typical size of mesoscopic devices, encouraging using graphene for electron quantum optics and quantum information processing.

In graphene, the quantum Hall (QH) effect reflects the Dirac Fermion properties of charge carriers, leading to a quantized conductance at $|\nu|G_0$, where $\nu = \pm 2, \pm 6, \pm 10, ...$ is the Landau level filling factor (positive and negative sign represents ν of electrons and holes, respectively) and $G_0 = e^2/h$ is the conductance quantum ($h$ is Planck's constant). Similar to QH states in standard two-dimensional electron systems, the conductance quantization can be explained by charge transport in edge channels that travel along the sample edge in one direction determined by the direction of the magnetic field $B$ and the carrier type. In a bipolar QH state, counter-circulating electron and hole edge modes merge and propagate in parallel in a PNJ (Fig. 1b). The conductance plateaus observed by experiments (ref: 7, 9) at $G_{\text{PNJ}} = G_0 |\nu_1||\nu_2|/(|\nu_1| + |\nu_2|)$ can be explained by assuming the full mode mixing in the PNJ (ref: 10, 11), where $\nu_1$ and $\nu_2$ are the filling factor in the *n* and *p* regions, respectively. However, experimental study of the mode mixing mechanism is lacking. If the mode mixing is caused by quasi-elastic scattering as suggested in (ref: 10, 11), a graphene PNJ acts as a beam splitter of electrons and holes. A better understanding of these properties is a crucial step towards the development of electron quantum optics experiments in graphene; beam splitters together with edge states are key components for electronic interferometry (ref: 12, 13, 14).

Shot noise measurements can provide insight into the mode mixing mechanism: when the electron and hole modes biased by $V_{sd}$ are mixed, the energy distribution in the PNJ $f_{\text{PNJ}}(E)$ becomes out-of-equilibrium and the subsequent partitioning of the modes gives rise to the shot noise. If the mode mixing is quasi-elastic, $f_{\text{PNJ}}(E)$ is a double-step function. At zero temperature, the shot noise generated by the partitioning of the modes with double-step $f_{\text{PNJ}}(E)$ is expected to be (ref: 10, Supplementary information),

$$S_I = 2eG_0 \frac{(|\nu_1||\nu_2|)^2}{(|\nu_1|+|\nu_2|)^3} V_{sd} = 2e \frac{|\nu_1||\nu_2|}{(|\nu_1|+|\nu_2|)^2} G_{\text{PNJ}} V_{sd}, \qquad (1)$$

characterized by the Fano factor $F = S_I/2eI$, yielding $F = 0.25$ for $(\nu_1, \nu_2) = (2, -2)$. Energy losses towards external degrees of freedom can drive $f_{\text{PNJ}}(E)$ towards a Fermi distribution with a chemical potential $eV_{sd}/2$ (Fig. 1**c**) (ref: 15, 16, 17), causing the noise (and thus the Fano factor) to vanish as the carrier dwell time in the PNJ becomes larger than the energy relaxation time (ref: 18). This dwell time can be tuned experimentally by changing the length of the PNJ. Inelastic processes between modes in the PNJ may occur, causing $f_{\text{PNJ}}(E)$ to relax towards a Fermi distribution with a finite temperature $T_{eff}(V_{sd})$ given by the balance between the Joule power dissipated in the PNJ and the heat flowing along the outgoing electronic channels (ref: 10, 19, Supplementary information). In this case, the Fano factor becomes $\tilde{F} = (3F)^{\frac{1}{2}}/\pi$ [$\tilde{F} \sim 0.28$ for $(\nu_1, \nu_2) = (-2, 2)$]. Note that standard transport measurements yield the same value of $G_{\text{PNJ}}$ for all cases, and thus cannot distinguish them.

In this work, we obtained bipolar graphene devices using a top gate covering half of the graphene (Fig. 1**a**); the carrier type in the gated region can be tuned by the gate voltage $V_G$, while that in the ungated region is fixed to electron by the doping (Methods). Therefore, the PNJ is formed at the interface between the gated and ungated regions when the carrier type in the gated region is hole for $\Delta V_G \equiv V_G - V_{\text{CNP}} < 0$ ($V_{\text{CNP}}$ is the gate voltage at the charge neutrality point). We prepared five samples with different interface lengths $L = 5, 10, 20, 50, 100$ μm. The direction of $B$ is chosen so that electron and hole modes from the ohmic contacts $C_{in}^e$ and $C_{in}^h$, respectively, merge at the PNJ. For the noise measurement, $V_{sd}$ is applied to either $C_{in}^e$ or $C_{in}^h$ and the noise is detected on $C_{det}$ (Fig. 1**a**) (Supplementary information). Magnetic fields up to $B = 16$ T have been applied. The base temperature is $T = 4.2$ K.

The inset of Fig. 2 shows the reflection of the averaged current from $C_{in}^e$ to $C_{det}$ in the sample with $L = 50$ μm. The magnetic field is $B = 10$ T, at which the filling factor in the ungated region is fixed at $\nu_{ug} = 2$. When the bipolar QH state at $(\nu_{ug}, \nu_g) = (2, -2)$ is formed for $\Delta V_G < -10$ V, the current injected from $C_{in}^e$ is partitioned equally to the electron and hole modes at the exit of the PNJ, yielding a reflection of 1/2. A current noise $S_I$ appears in this regime. As $V_{sd}$ applied to $C_{in}^e$ is increased, the excess noise $\Delta S_I \equiv S_I - S_I(V_{sd} = 0)$ increases (solid cyan circles in Fig. 2). $\Delta S_I$ approaches linear behavior for $eV_{sd} > k_B T$, characteristic of the shot noise. A similar signal appears when $V_{sd}$ is applied to $C_{in}^h$ (open cyan circles). In the unipolar QH state at $(\nu_{ug}, \nu_g) = (2, 2)$ for $20 < \Delta V_G < 50$ V, on the other hand, the shot noise is zero (solid black circles), proving that the shot noise is indeed generated at the PNJ.

Quantitatively, we extracted $F$ by fitting $\Delta S_I$ as a function of $V_{sd}$ using the relation including temperature broadening (ref: 20):

$$\Delta S_I = 2eFG_{\text{PNJ}}V_{sd}\left[\coth\left(\frac{eV_{sd}}{2k_BT}\right) - \frac{2k_BT}{eV_{sd}}\right], \tag{2}$$

where $G_{\text{PNJ}}$ is obtained by average current measurements. The fit yields $F = 0.018$, which is one order of magnitude smaller than $F = 0.25$ expected for the noise from the double-step energy distribution. This indicates that $f_{\text{PNJ}}(E)$ evolves during the charge propagation for $L = 50$ μm, reducing the shot noise.

The evolution of $f_{\text{PNJ}}(E)$ can be investigated using samples with different $L$. Figure 3**a** shows the results of the noise measurement in the bipolar QH state at $(\nu_{ug}, \nu_g) = (2, -2)$ for the five samples with $L$ between 5 and 100 μm. The data show that the shot noise decreases with increasing $L$ and almost disappears at $L = 100$ μm (Fig. 3**b**), indicating that $f_{\text{PNJ}}(E)$ relaxes to the thermal equilibrium through interactions with external degrees of freedom. An exponential fit of the data yields a relaxation length $L_0 = 16$ μm. The extrapolation to $L = 0$ gives $F \sim 0.27$, consistent with the limit of quasi-elastic scattering $F = 0.25$. Furthermore, the decrease is well reproduced by a model gradually coupling the modes propagating in the PNJ to cold external states (Supplementary information). Note that we are not able to observe whether inelastic scattering occurs inside the PNJ, because of the large error bars explained below. An important implication of the results is that, within the typical scale of usual mesoscopic devices (< 1 μm), the energy loss towards external degrees of freedom is negligible and the current channels in the PNJ can be regarded as an isolated system.

We further investigate the properties of the PNJ focusing on the energy relaxation mechanism by measuring the shot noise for a wide range of $B$ and $\Delta V_G$. We identify the electronic states in the gated and ungated regions as a function of $B$ and $\Delta V_G$ by a low-frequency current measurement from $C_{in}^e$ to $C_{det}$ (Fig. 4a) and then investigate the relation between those states and the noise. The electronic state in the ungated region depends only on $B$ and the $\nu_{ug} = 2$ QH state is formed for $B > 4$ T. In the gated region, the non-QH states at $\nu_g = 8, 4, 0,$ and $-4$ appear as a current peaks. The bipolar QH state at $(\nu_{ug}, \nu_g) = (2, -2)$ is formed for $B > 4$ T and between $\nu_g = 0$ and $-4$ (the region indicated by dashed lines), in which the current is almost constant, consistent with the quantized conductance (ref: 7, 8, 9). The shot noise in the sample with $L = 10$ μm becomes small (Fig. 4b) when either or both ungated and gated regions are in a non-QH state. This confirms that shot noise is generated by the PNJ in a well-developed bipolar QH state. Within the bipolar QH state, the shot noise fluctuates largely, depending on $B$ and $\Delta V_G$. This noise fluctuation cannot be ascribed to $G_{\text{PNJ}}$, which is almost constant in the bipolar QH state. Furthermore, since the noise is generated in the well-developed QH state, the existence of multiple noise sources is unlikely. These facts indicate that the noise fluctuation is due to the fluctuation of the energy relaxation rate, which induces the fluctuation of Fano factor. Fig. 4c shows the histogram of the Fano factor in the bipolar QH state calculated using equation (2). The standard deviation is about 50% of the mean value.

The random variation of the energy relaxation rate as a function of $B$ and $\Delta V_G$ suggest that localized states in bulk graphene play a main role for the energy relaxation. Energy in the PNJ can escape to the bulk graphene through Coulomb interaction with localized states: high frequency potential fluctuations in the PNJ, which is the source of the shot noise, are dissipated in the localized states. Since the energy level and the profile of the localized states depend on $B$ and $\Delta V_G$, fluctuations of the relaxation rate can be induced. On the other hand, the average current through the PNJ, which merely reflects the transmission coefficient, is hardly affected by the localized states. Note that a simple model of interaction with 2D phonons in the PNJ fails to quantitatively reproduce our observations (Supplementary information). It is reported that the electron-phonon coupling is expected to be vanishingly small in usual unipolar edge channels (ref: 21, 22, 23). To understand the energy relaxation length quantitatively, detailed analysis including interactions with phonons and any other possible mechanisms for the energy relaxation is necessary.

In conclusion, we showed that the mode mixing at PNJ in graphene bipolar QH states leads to non-equilibrium $f_{\text{PNJ}}(E)$, generating shot noise. For a short PNJ ($L \ll 16$ μm), the energy loss towards external states is negligible and the noise is consistent with a quasi-elastic mode mixing. This suggests that a graphene PNJ can act as a beam splitter. Since 16 μm is much larger than typical length scale of mesoscopic devices, our results encourage using graphene for electron quantum optics experiments and quantum information.

**Method**

**Device fabrication.** We prepared a graphene wafer by thermal decomposition of a 6H-SiC(0001) substrate. SiC substrates were annealed at around 1800 °C in Ar at a pressure of less than 100 Torr. For the fabrication of devices, graphene was etched in an $O_2$ atmosphere. After the etching, the surface was covered with 100-nm-thick hydrogen silsesquioxane (HSQ) and 60-nm-thick $SiO_2$ insulating layers. As a result of doping from the SiC substrate and the HSQ layer, graphene has n-type carriers with the density of about $5 \times 10^{11}$ cm$^2$. The width of the PNJ is roughly corresponds to the thickness of the insulating layers and estimated to be 200 nm at most. In the QH effect regime, because of the Landau level quantization, the width becomes smaller with *B*. An important advantage of the SiC graphene is its size: it is single domain for 1 cm$^2$, allowing us to investigate the effect of PNJ length.

**Noise measurement.** For the noise measurement, the current noise is converted into voltage fluctuations across one 2.5 kΩ resistor in series with the sample. A 500 kHz bandwidth 3 MHz tank circuit combined with a home-made cryogenic amplifier is used. Accurate calibration of the noise is done using Johnson-Nyquist noise that relies on the quantification of the resistance at $\nu = 2$ and the temperature of the system.

**Acknowledgements**

The authors acknowledge funding from the ERC Advanced Grant 228273 MeQuaNo and the ANR MetroGraph grant. The authors are grateful to S. Tanabe, P. Jacques, and M. Ueki for experimental support.

**Author contributions**

N. K. and P. R. performed experiments. N. K., F. D. P., P. R., and D. C. G. analyzed the data and wrote the manuscript. H. H. grew the wafer.

**Additional information**

Supplementary information is available.

**Competing interests statement**

The authors declare that they have no competing financial interests.

**Figure 1 | Schematic of the device. a**, The top gate covers the right half of the graphene. In the ungated region, the carrier type is electron and the density is fixed at about $5 \times 10^{11}$ cm$^{-2}$. In the gated region, the carrier type can be changed to hole by applying negative gate voltage $V_G$. A source-drain bias $V_{sd}$ is applied to either $C_{in}^e$ or $C_{in}^h$ and the noise is measured on $C_{det}$. The upperright contact is comb-shaped to achieve a good contact to the $p$ region (supplementary information). **b**, QH edge modes for the unipolar (left) and bipolar regimes (right). **c**, $f_{\text{PNJ}}(E)$ in the quasi-elastic case (left). In the presence of energy relaxation towards external degrees of freedom, $f_{\text{PNJ}}(E)$ becomes a Fermi distribution at base temperature (right). Additional inelastic scattering between modes in the PNJ may drive towards a Fermi distribution at finite temperature $T_{eff}$ (bottom), which can then relax towards a cold Fermi distribution.

**Figure 2 | Shot noise generated by *p-n* junction.** Excess noise $\Delta S_I$ of the sample with $L = 50$ μm as a function of $V_{sd}$ for the bipolar QH state at $(\nu_{ug}, \nu_g) = (2, -2)$ (cyan circles) and the unipolar QH state at $(\nu_{ug}, \nu_g) = (2, 2)$ (black circles). For the solid and open circles, $V_{sd}$ is applied to $C_{in}^e$ and $C_{in}^h$, respectively. The solid trace is the result of a fit using Eq. (2). Inset: reflection of the average current from $C_{in}^e$ to $C_{det}$ as a function of $\Delta V_G$. The cyan and black circles represent the $\Delta V_G$ at which the noise is measured.

**Figure 3 | Shot noise as a function of *p-n* junction length. a**, $\Delta S_I$ in the bipolar QH state at $(\nu_{ug}, \nu_g) = (2, -2)$ as a function of $V_{sd}$ for the five samples with $L$ between 5 and 100 μm. All data are taken at $B = 10$ T. Lines are results of the fit using Eq. (2), by which the values of $F$ (indicated on the right-hand side of the figure) are obtained. **b**, $F$ as a function of $L$. The error bar represents the fluctuations of the extracted $F$ for different values of $B$ and $\Delta V_G$ in the bipolar QH state (see Fig. 4c for the details). An exponential fit (blue curve) yields a relaxation length $L_0 = 16$ μm. The dashed horizontal line represents the expected value $F = 0.25$ for the double-step $f_{\text{PNJ}}(E)$.

**Figure 4 | Fluctuations of shot noise. a**, Gray scale plot of the current from $C_{in}^e$ to $C_{det}$ at 2 kHz for the sample with $L = 10$ μm. The current is measured through the resonator and the amplifier. Black dashed lines indicate the region for the bipolar QH state at $(\nu_{ug}, \nu_g) = (2, -2)$. Red dotted lines represent the non-QH state in the gated region at $\nu_g = -4, 0, 4$ and 8. **b**, Color scale plot of $\Delta S_I$ for the sample with $L = 10$ μm as a function of $\Delta V_G$ and $B$. The applied current is fixed at 400 nA, which corresponds to $V_{sd} = 5.2$ mV in the bipolar QH state at $(\nu_{ug}, \nu_g) = (2, -2)$. The black dashed and red dotted lines correspond to those in **a**. **c**, Histogram of the Fano factor within the bipolar QH state. The red dot and bar represent the mean value and the standard deviation, which correspond to the data point and the error bar in Fig. 3**b**, respectively.

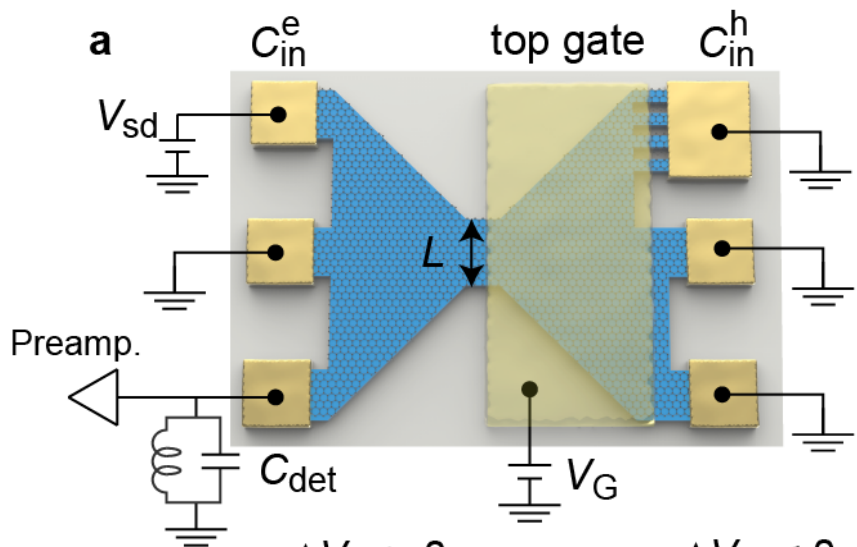
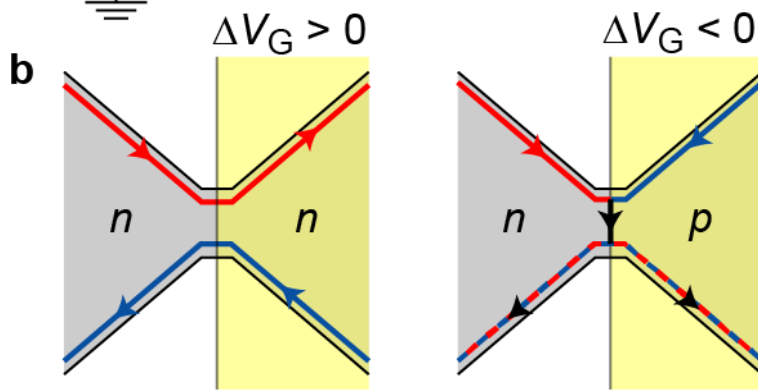
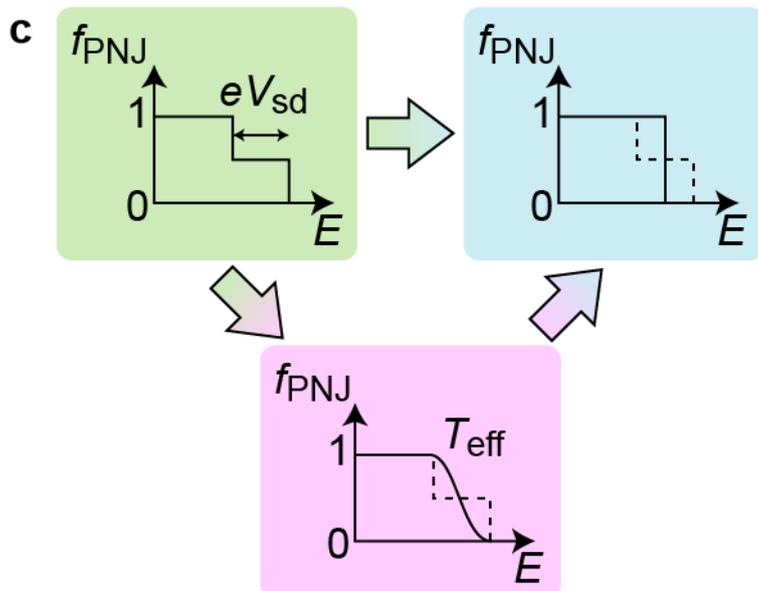

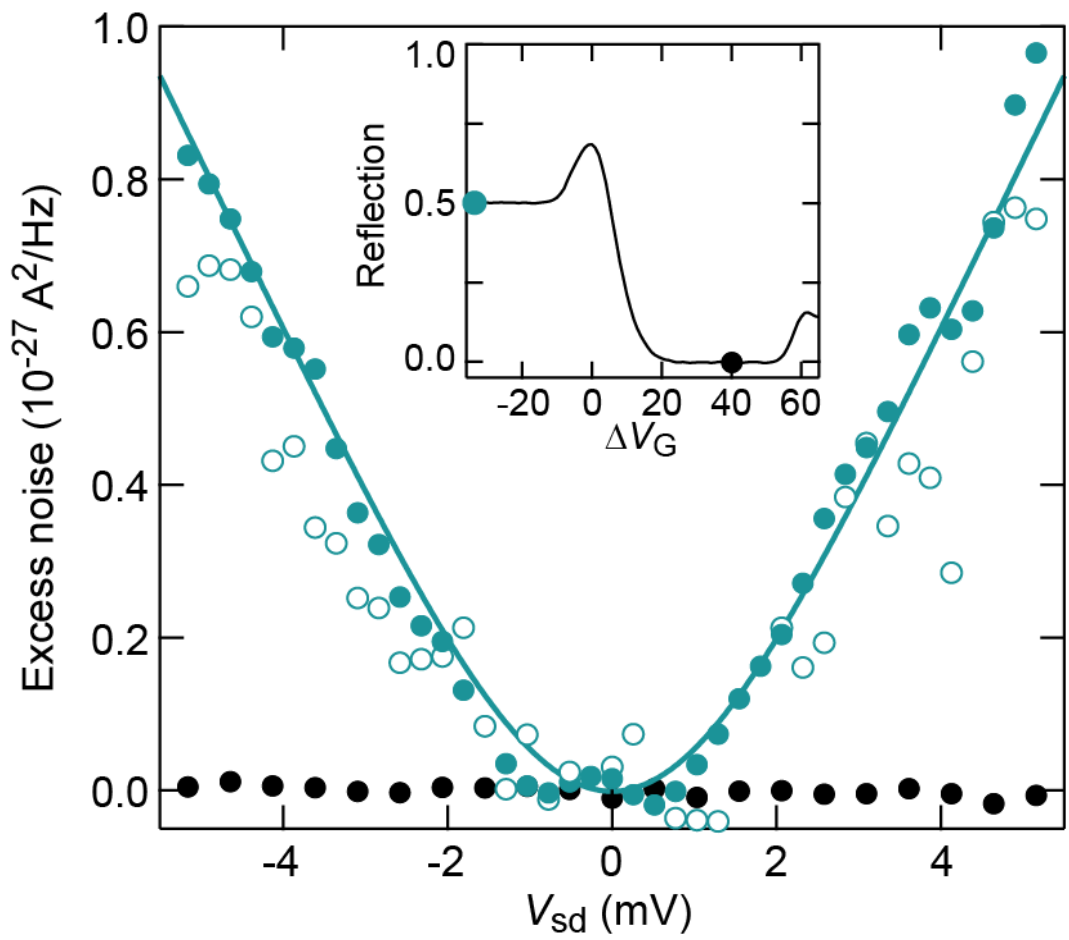

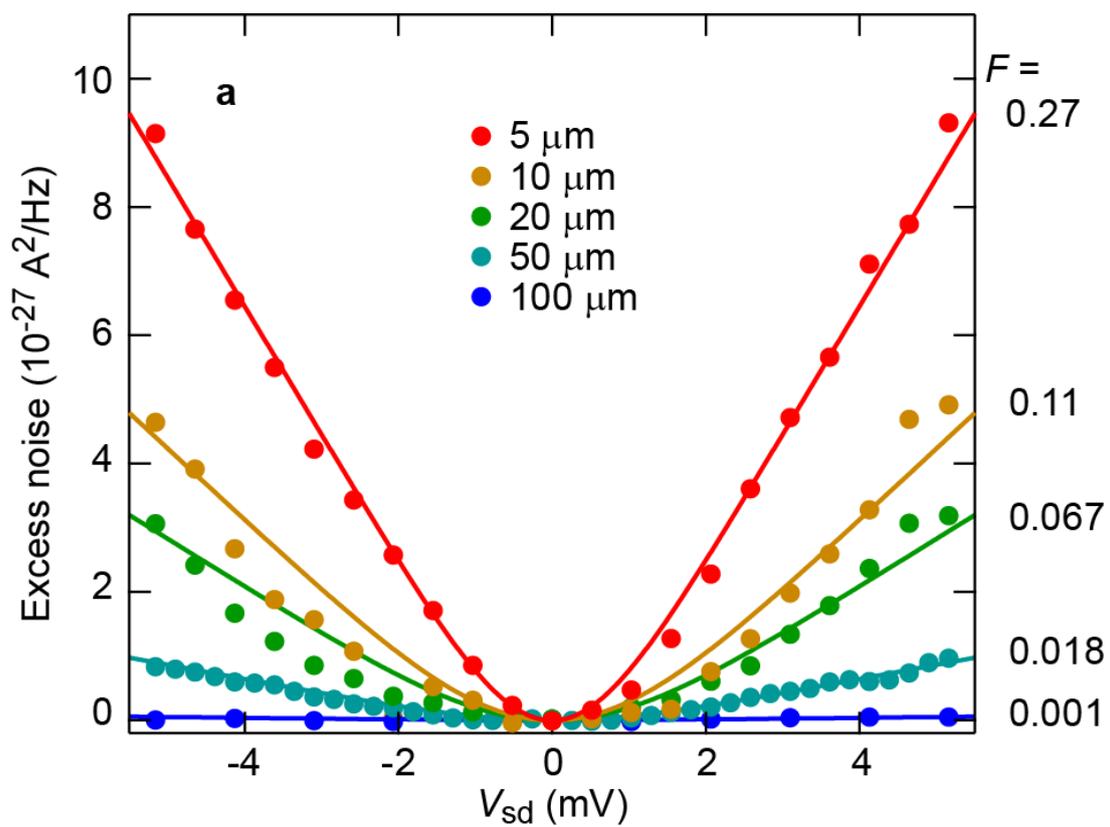

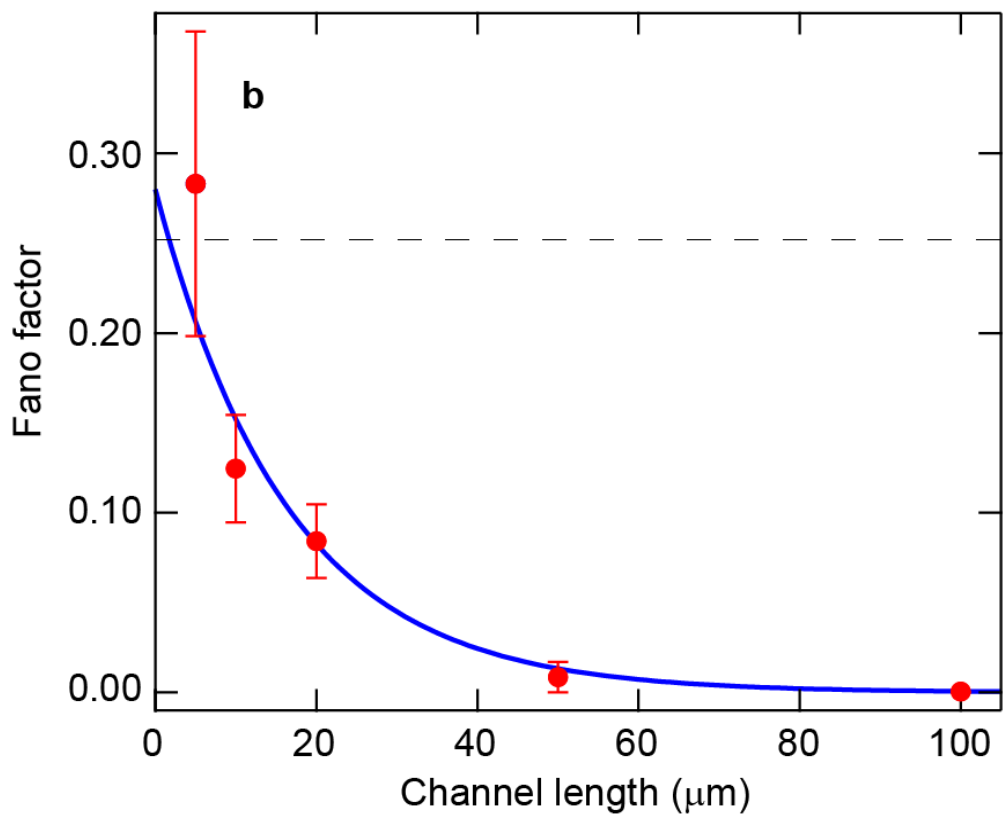

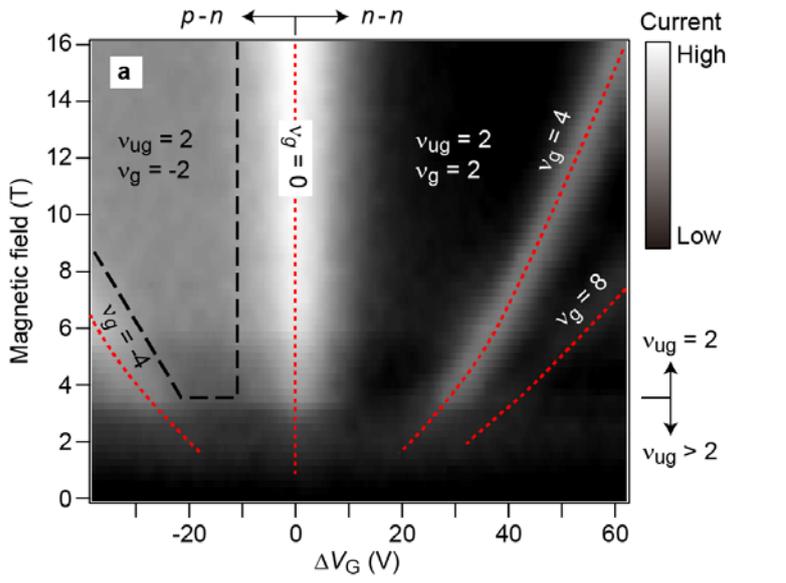

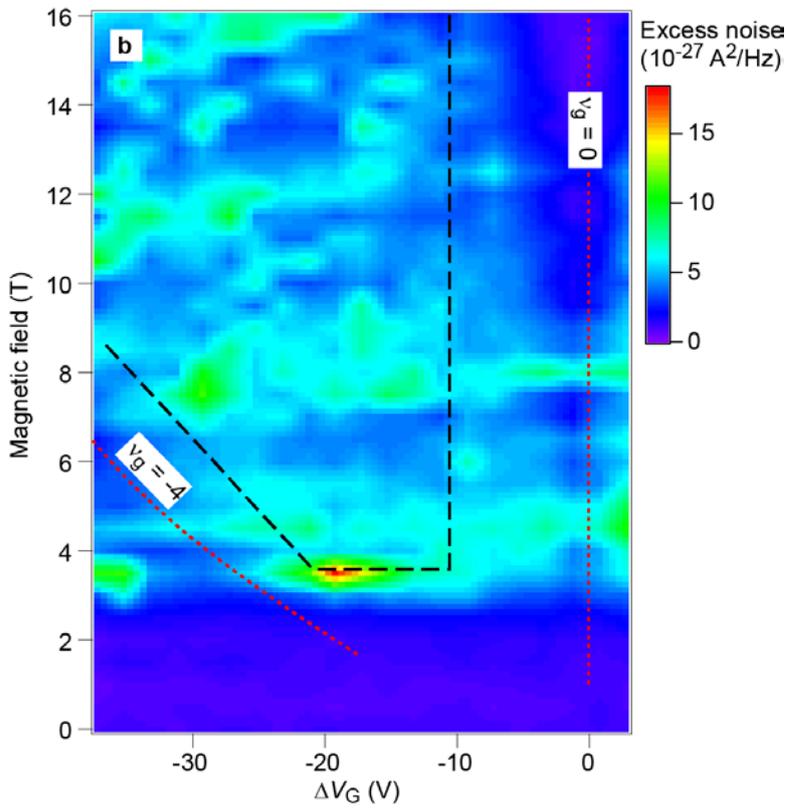

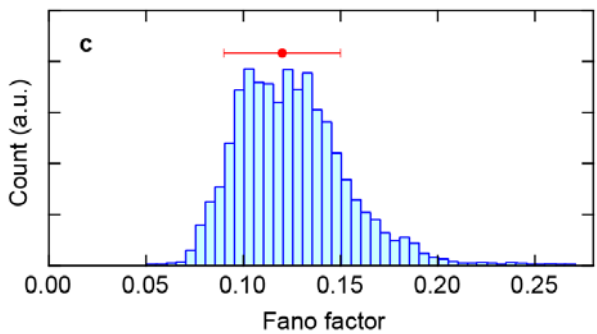